\documentclass[journal]{IEEEtran}

\usepackage{amssymb,amsmath}
\usepackage{graphicx}
\usepackage{epsfig}
\usepackage{amsfonts}
\usepackage{rotating}
\usepackage{amssymb}
\usepackage{amsbsy}
\usepackage{amsmath}
\usepackage{times}
\usepackage{algorithmic}
\usepackage{algorithm}
\usepackage{openbib}
\usepackage[noadjust]{cite} 
\usepackage{flushend}
\usepackage{enumerate}
\usepackage{color}
\usepackage{psfrag,fancyhdr}
\usepackage{acronym}
\usepackage{hhline}
\usepackage{latexsym}
\usepackage{multirow}
\usepackage{float}
\usepackage[english]{babel}
\usepackage{url}
\usepackage{algorithm,algorithmic}
\usepackage{setspace}
\usepackage{xspace}
\usepackage{supertabular}
\usepackage{eucal} 
\usepackage{blindtext}
\usepackage{scrextend}
\usepackage{longtable}
\usepackage[normalem]{ulem}
\addtokomafont{labelinglabel}{}

\usepackage{listings}

\usepackage{mathrsfs}
\usepackage{soul,xcolor}
\usepackage{caption}
\usepackage{subcaption}
\usepackage{makecell} 
\usepackage{hyperref} 
\usepackage{orcidlink}
\newcommand{\naive}{Na\"ive }

\definecolor{lightblue}{rgb}{.4,1,1}
\newcommand{\KSM}[1]{
}
\definecolor{lightpink}{rgb}{1,.9,.9}

\usepackage{booktabs}

\usepackage[flushleft]{threeparttable}
\usepackage{tikz}
\usetikzlibrary{shapes}
\usetikzlibrary{plotmarks}

\begin{document}

\title{Advancing SQL Injection Detection for High-Speed Data Centers: A Novel Approach Using Cascaded NLP}

\author{Kasim~Tasdemir \orcidlink{0000-0003-4542-2728}, 
Rafiullah Khan \orcidlink{0000-0003-1478-1568}, 
Fahad~Siddiqui \orcidlink{0000-0002-4334-9478},~\IEEEmembership{Member,~IEEE}, 
Sakir~Sezer \orcidlink{0000-0003-2857-616X},~\IEEEmembership{Member,~IEEE}, 
Fatih~Kurugollu \orcidlink{0000-0002-2508-4496},~\IEEEmembership{Member,~IEEE},  
Sena Busra Yengec-Tasdemir \orcidlink{0000-0001-8322-4832},
Alperen~Bolat \orcidlink{0000-0003-0170-8038}

\thanks{K. Tasdemir, S. Sezer, {S.B. Yengec-Tasdemir}, A. Bolat are with \textit{Centre for Secure Information Technologies}, Queen's University Belfast, Belfast, United Kingdom. They are members of GDR-LAB group at QUB (http://www.gdr-lab.com). Email: \{k.tasdemir,  s.sezer, s.yengectasdemir, a.bolat\}@qub.ac.uk
}
\thanks{R. Khan and F. Siddiqui are with \textit{NVIDIA Corporation}, Belfast, United Kingdom. Email: \{rafiullahk, fsiddiqui \}@nvidia.com,
}
\thanks{F. Kurugollu is with \textit{Department of Computer Science, College of Computing and Informatics}, 
University of Sharjah, Email: fkurugollu@sharjah.ac.ae }

\thanks{This project is funded as a part of Grant for R\&D by Invest NI (Grant: RD08201502)}
\thanks{Note: This work has been submitted to the IEEE for possible publication. Copyright may be transferred without notice, after which this version may no longer be accessible.}

}

\maketitle

\begin{abstract}

Detecting SQL Injection (SQLi) attacks is crucial for web-based data center security, but it's challenging to balance accuracy and computational efficiency, especially in high-speed networks. Traditional methods struggle with this balance, while NLP-based approaches, although accurate, are computationally intensive.

We introduce a novel cascade SQLi detection method, blending classical and transformer-based NLP models, achieving a 99.86\% detection accuracy with significantly lower computational demands—20 times faster than using transformer-based models alone. Our approach is tested in a realistic setting and compared with 35 other methods, including Machine Learning-based and transformer models like BERT, on a dataset of over 30,000 SQL sentences.

Our results show that this hybrid method effectively detects SQLi in high-traffic environments, offering efficient and accurate protection against SQLi vulnerabilities with computational efficiency. The code is available at \href{https://github.com/gdrlab/cascaded-sqli-detection}{GitHub}\footnote{https://github.com/gdrlab/cascaded-sqli-detection}.

\end{abstract}


\IEEEpeerreviewmaketitle

\section{Introduction}
\label{sec:Introduction}

SQL injection is a prevalent type of cyber attack that poses a serious threat to the security of web applications and data centers \cite{owasp2021}. In an SQL injection attack, a malicious user exploits vulnerabilities in the application's input validation mechanisms to execute arbitrary SQL statements on the database, potentially allowing them to access or modify sensitive data \cite{defendingAgainst,WAFReinforcement}.

The consequences of a successful SQL injection attack can be severe, ranging from unauthorised access to sensitive information to the complete compromise of the underlying system. As a result, detecting and mitigating SQL injection attacks is a critical task for any organisation that relies on web applications and databases to store and process sensitive data.

The traditional SQLi detection is running a rule-based pattern matching tests on the payloads. However, this approach has serious limitations, especially when performing analysis on complex SQL queries which might look legitimate on the surface. This created a need for analysing the queries at higher levels. To alleviate this problem, Natural Language Processing (NLP) based SQL injection detection has been proposed by several previous studies \cite{M2022, Hatomura2021, Ahmed2020, Triloka2022, Gogoi2021, deepsqli, Matam2022, Ghozali2022, Li2019, Zhang2022, Lakhani2022, Seyyar2022, K2022, Oudah2023}. 

NLP-based SQL injection detection is a more effective and accurate approach than traditional pattern-matching-based approaches for several reasons.

\begin{figure}
	\centering
	\includegraphics[width=0.5\textwidth]{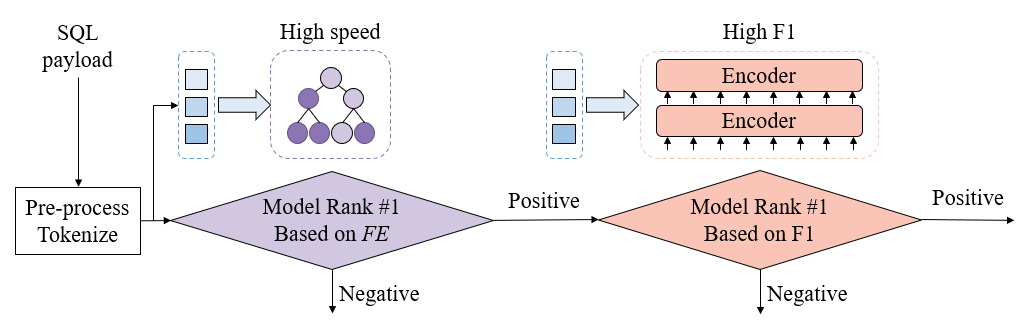}
	\caption{The high-level structure of the proposed cascade SQLi attack detection method, as inspired by Viola et al.~\cite{viola2001rapid}, is illustrated. In the first stage, the method effectively captures over 99.73\% of potential attacks (recall rate) with a significantly low computational burden, thereby enabling efficient detection of SQLi attacks. Subsequently, the suspicious SQL payloads are passed to the second stage for further investigation. The transformer-based model employed in the second stage conducts a re-analysis of the SQL payloads, thereby mitigating possible false alarms. This two-stage cascade system enhances the overall detection speed by a factor of 20\texttimes, while maintaining a high detection accuracy, as demonstrated by an F1 score of 0.9981. Notably, the model utilised in the first stage can be dynamically replaced based on the introduced \emph{FE} score, thereby enabling adaptive and optimised detection performance. 
 }
	\label{fig:two_stage}
\end{figure}

First, traditional pattern-matching approaches rely on predefined patterns and rules to detect SQL injection attacks. These patterns are often static and inflexible, which can result in high false positive rates and an inability to detect novel attack patterns. In contrast, NLP-based approaches use machine learning techniques to learn from large amounts of data and dynamically adapt to new attack patterns, leading to more accurate and effective detection.

\begin{figure*}[th]
     \centering
     \begin{subfigure}{0.49\textwidth}
         \centering
         \includegraphics[width=\textwidth]{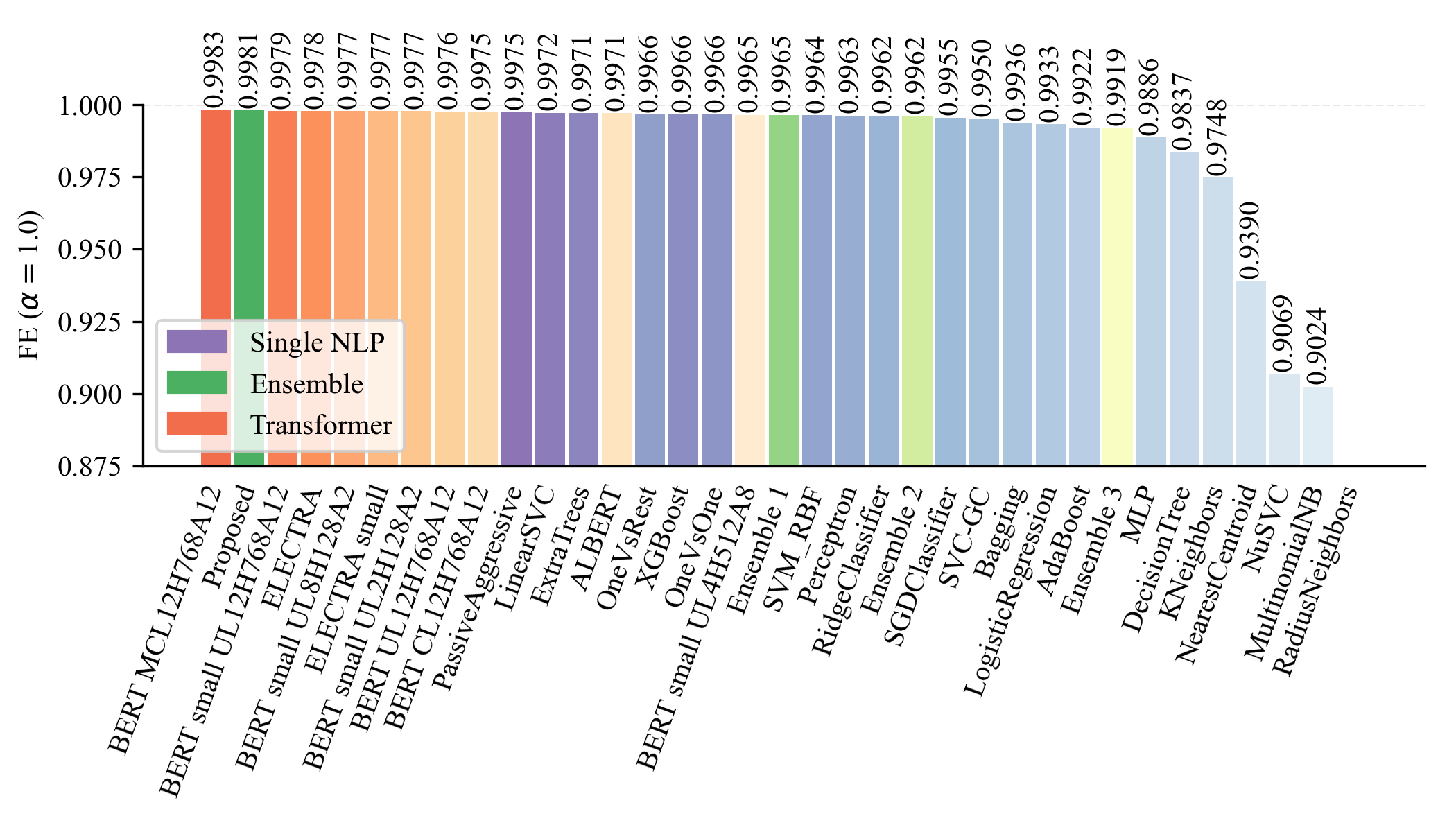}
         \caption{}
         \label{fig:sub:FE-100-all}
     \end{subfigure}
     \hfill
     \begin{subfigure}{0.49\textwidth}
         \centering
         \includegraphics[width=\textwidth]{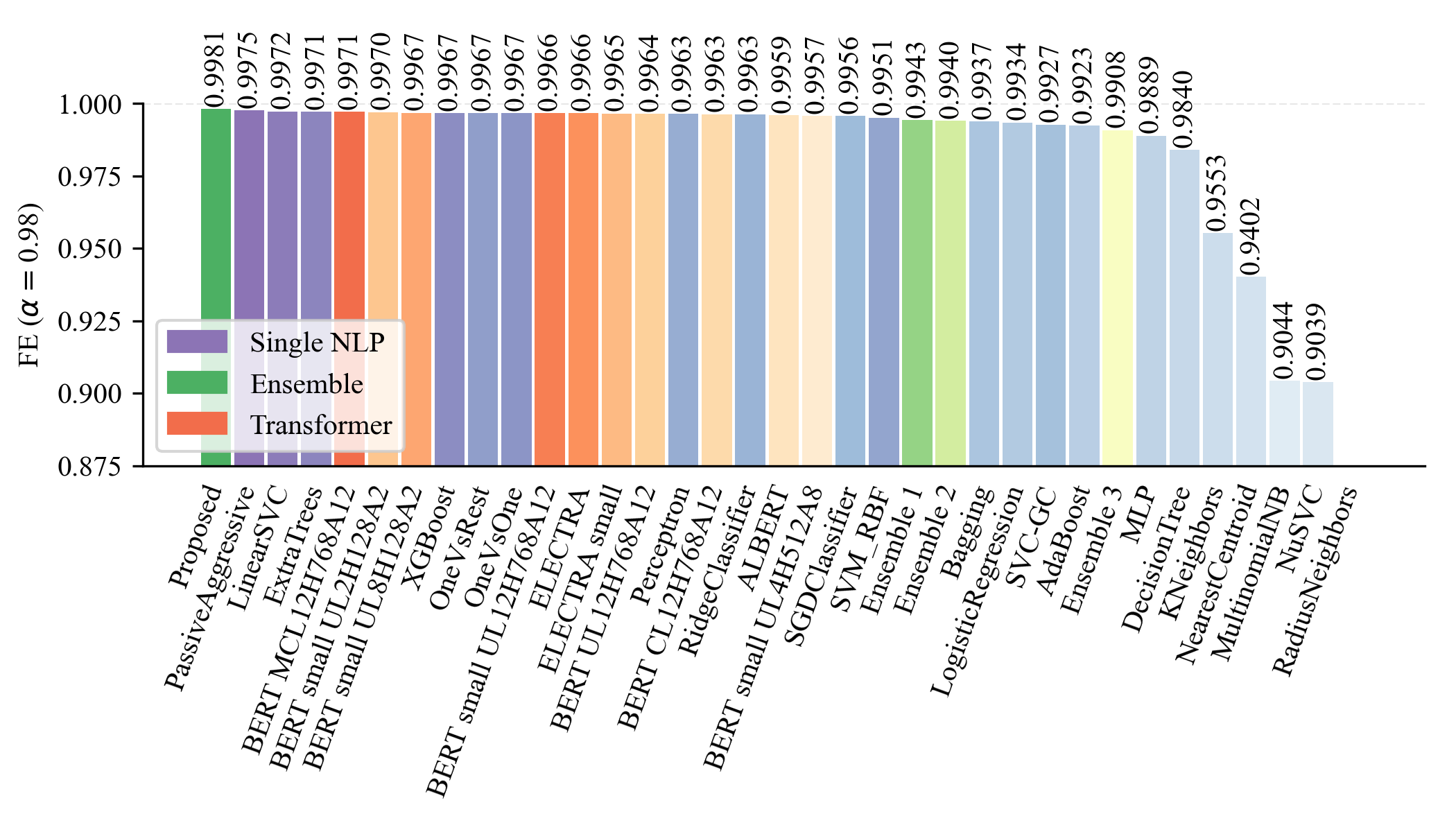}
         \caption{}
         \label{fig:sub:FE-90-all}
     \end{subfigure}
        \caption{
        A comparative analysis of all methods is conducted based on the introduced F1 efficiency (FE) score, considering two scenarios: (a) when $\alpha=1.0$, and (b) when $\alpha=0.98$. The findings reveal that the proposed method emerges as the most favourable choice when prioritising 98\%-2\% importance to the F1 score and inference latency, respectively. When the FE score is adjusted from 1.00 to 0.98, the Transformer-based models shift towards the lower end of the list due to their higher inference latencies. In scenarios where computational resources are limited and emphasis is placed on speed and accuracy (i.e., 2\% and 98\% importance, respectively), the proposed method demonstrates superior performance. The method categories and their order within the group, as denoted by colours and shades, respectively, are consistent with those presented in Fig.~\ref{fig:all-F1-vs-inference}.
        }
        \label{fig:FE-all}
\end{figure*}

Another reason is, NLP-based approaches can analyse the contextual meaning of the input data and identify anomalies in the input that is indicative of an SQL injection attack. This is in contrast to traditional pattern-matching approaches, which only look for specific patterns of characters and symbols that match known SQL injection attack strings. As a result, NLP-based approaches can detect more subtle and sophisticated SQL injection attacks that may bypass traditional pattern-matching methods.

NLP-based approaches can analyse the intent behind the input data and identify potentially malicious user behaviour, such as attempts to bypass input validation mechanisms or execute unauthorised SQL queries. This is not possible with traditional pattern-matching approaches, which only focus on detecting specific patterns of characters and symbols.

In summary, NLP-based SQL injection detection is a more effective and accurate approach than traditional pattern matching-based approaches due to its ability to dynamically adapt to new attack patterns, analyse the contextual meaning of input data, identify malicious user behaviour, and leverage contextual information to improve detection accuracy.

However, the scale of network traffic on modern data centers and DPUs makes it challenging to detect SQL injection attacks efficiently and quickly using computationally demanding Transformer-based detection methods only. Considering Nvidia Bluefield 3 DPU supporting network traffic at 400Gbits bandwidth,  at that scale, the NLP-based SQLi detection techniques can be computationally intensive and lead to significant performance overhead and delays. 

To address these challenges, there is a growing need for more efficient and accurate SQL injection detection techniques that can be seamlessly integrated into the complex infrastructure of data centers and DPUs. These techniques should be designed to perform higher-level analysis of SQLi queries and leverage the capabilities of specialized hardware, such as Nvidia Bluefield 3, which can provide high-performance computing and networking capabilities to support real-time detection and mitigation of SQL injection attacks.

In this manuscript, we introduce a novel approach for SQLi detection that is specifically designed to prioritize high efficiency and accuracy. Our approach leverages a combination of fast classical machine learning-based NLP techniques to enable high detection speed, along with recent Transformer-based NLP methods to achieve superior accuracies while minimizing false negatives (missed detections). We comprehensively evaluate the performance of our approach using a publicly available SQLi dataset and demonstrate its effectiveness in detecting SQL injection attacks. Our results substantiate that our approach presents a reliable and efficient solution for SQLi detection in data centers and DPUs, thereby enhancing the security of web applications and databases against cyber threats.

The key contributions and novelties introduced in this paper are:

\begin{itemize}
\item The proposition of a unique cascading SQLi detection model that seamlessly blends classical Machine Learning (ML) classifiers and cutting-edge transformer-based NLP techniques. This novel approach has been designed to effectively and accurately detect malicious queries, offering comparable detection accuracy at higher speeds, a combination previously unexplored,
\item The introduction and implementation of a novel performance metric, F1 Efficiency. This measure offers an objective way of comparing different methodologies based on user preferences for the speed-accuracy trade-off (see Eq.\ref{eq:f1-fe}). It also allows for the dynamic alteration of the first stage model when there's a need to reduce the computational load on the host system,
\item A detailed comparison of 35 different methods, inclusive of both ML classifiers and contemporary transformer-based models, with an evaluation of their classification and speed performance (see Section\ref{sec:experiments}). This exhaustive comparison adds value to the understanding of performance dynamics across a diverse range of methodologies,
\item The integration and evaluation of various ensemble models that combine classic NLP features and ML models in the tests, contributing to a richer and more diverse experimental setup.
\end{itemize}

The organisation of the paper is as follows. In Section~\ref{sec:BackgroundRelatedWork}, the overview of the related previous studies is presented. The following Section~\ref{sec:NLPDetectionOfSQLInjectionAttack} gives technical descriptions of the NLP application on SQLi detection. Section~\ref{sec:proposed} explains the details of the proposed method. The experimental results are demonstrated in Section~\ref{sec:experiments} and finally, the conclusion of the paper is given in Section~\ref{sec:Conclusion}.

\section{Background \& Related Work}
\label{sec:BackgroundRelatedWork}

Current SQLi attack detection methods can be broadly classified into three categories.

The first approach, which is the traditional one, employs rule-based pattern matching on SQL queries. This method is widely used in industrial solutions. However, it has several limitations, such as being static and inflexible, which can lead to high false positive rates and an inability to detect novel attack patterns.

The research community has attempted to address these limitations by employing machine learning techniques, which constitute the second category of SQLi detection methods. These methods are typically based on hand-crafted features and natural language processing (NLP) techniques. While they offer improved accuracy over traditional rule-based methods, they are still limited by their dependence on pre-defined features and their inability to adapt to new attack patterns.

With the advent of transformer-based NLP models, a limited number of attempts have been made to incorporate them in SQLi detection tasks, which constitute the third category of methods. Transformer-based approaches have shown promising results in detecting SQLi attacks, achieving high detection accuracy in exchange for considerable computation and memory loads.

In the literature, there is a limited number of studies using transformer-based models targeting SQLi detection. One related work, DeepSQLi \cite{deepsqli}, attempts to solve the automatic testing SQLi vulnerabilities of web services. The used model is based on a vanilla self-attention transformer. However, the aim of the study is not to detect malicious SQL statements, it is to evaluate the vulnerabilities of web applications against such attacks. Another study employs BERT against several cyberattacks \cite{Matam2022}. They first recursively decode the user input from encoded HTTP payloads into raw texts. The text is fed to the BERT for classification. Similar approaches are proposed in \cite{Lakhani2022, Seyyar2022, K2022}. 

The classical ML-based NLP methods have been investigated in previous studies. In \cite{Oudah2023}, the authors propose an SQLi detection system comprised of TF-IDF features and SVM, XGBoost-based and Na\"ive Bayes classification parts. They tokenise the content with BoW, BoC and n-gram sequence lengths. Similarly, TF-IDF is also used in \cite{Ghozali2022, Li2019, Zhang2022} for SQLi detection. In \cite{Tasdemir23}, the speed and detection performance of classical ML models on DPU for SQLi detection task is investigated.

Apart from those, there are other NLP based SQLi detection methods proposed in the literature \cite{M2022, Hatomura2021, Ahmed2020, Triloka2022, Gogoi2021}.

The majority of the previous studies target vulnerability detection of web applications. The efficiency and practical feasibility of the methods are not the main concerns. However, this study is aimed at designing a system with high detection accuracy and low computational cost.

\section{NLP for Detection of SQL Injection Attack}
\label{sec:NLPDetectionOfSQLInjectionAttack}

\begin{figure}
	\centering
	\includegraphics[scale=0.5]{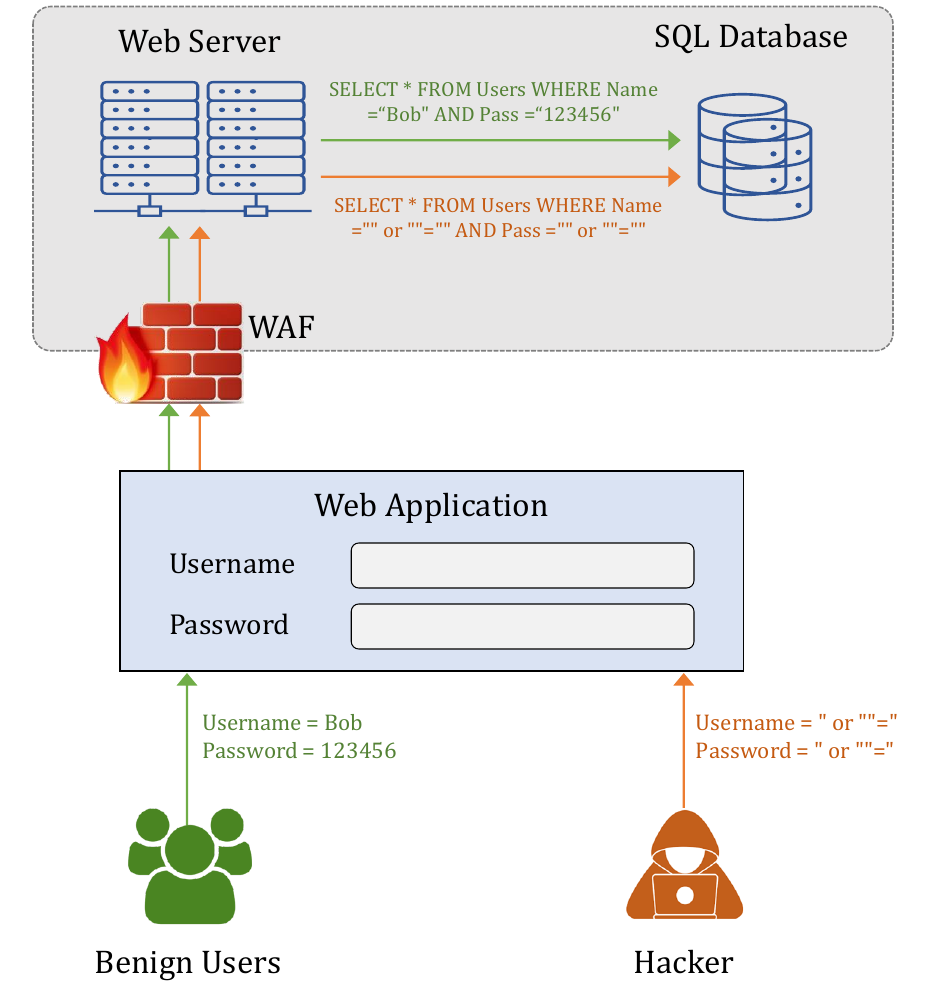}
	\caption{SQL injection attack. Benign users provide their original usernames and passwords. A hacker will try SQL injection statements such as ``or" ``=" that is always TRUE and will retrieve all rows from the `Users' table in the database.}
	\label{fig:sqli_concept}
\end{figure}

SQL is a declarative language which has high-level abstractions of procedures. As in all high-level languages, the same operation can be generated using different statements. For example, one can create a SQL tautology in infinite ways, i.e., \lstinline[columns=fixed,language=SQL, breaklines]{'abc'='abc' OR 1=1 OR 99=99}... Therefore, it is not possible to provide all possible SQL injection statements to a classical pattern-matching-based detection method. Even if these statements are not exactly the same, their higher-level meanings (latent representations) are similar in a way. At this point, NLP algorithms can help us to extract the features from a latent space where the representations of SQLi statements reside close to each other. Unlike the classical pattern matching approaches, we can therefore create task-specific latent spaces to effectively classify malicious queries. This is the main motivation behind using NLP algorithms on SQL statements.

The main objective of the proposed method is to detect SQL injection queries efficiently without compromising SQLi detection accuracy. Even if there are several approaches that can be used for performing NLP analysis on the queries, classical methods such as Latent Semantic Analysis (LSA) become stronger candidates than Transformer or Deep neural architectures when computational efficiency is a major concern. LSA approach requires extraction of Document-term matrix (DTM) and application of Singular Value Decompositon (SVD) for dimensionality reduction. For the sake of computational efficiency, the SVD step is replaced with Term Frequency Inverse Document Frequency (TF-IDF) and machine learning based classification in the ensemble algorithms.
\subsection{TF-IDF, BoC, BoW}
\label{sec:sub:tfidf}

DTM is an occurrence matrix, $F$, whose elements are multiplicities of the terms, $t$, in a set of documents, $D$. Its $d$\textsuperscript{th} row $t$\textsuperscript{th} column is denoted as $f_{t,d}$. Usually, $F$ is processed by TF-IDF before further analysis.

TF-IDF is a statistical NLP feature commonly used in LSA, information retrieval, text and sentiment analysis. It indicates the relevance of the representations of the \textit{terms}, $t$, in a corpus, $D$. If a term appears quite often in a document under test, it is reasonable to assume a relationship between the frequency of the term and the document. 

One way to measure the Term Frequency (TF) is to normalise the raw number of occurrences (\textit{multiplicity}) $f_{t,d}$ of the term, $t$, in a  specific document $d\in D$ as given in Eq.~\ref{eq:tf}.   

\begin{equation}
    \mathrm{\overline{F}}=[\bar{f}_{i,j}], \bar{f}_{t,d} = \frac{f_{t,d}}{{\sum_{t' \in d}{f_{t',d}}}}
    \label{eq:tf}
\end{equation}

\begin{equation}
    df(t) = |\{d\in D | t\in d\}|
    \label{eq:df}
\end{equation}

On the other hand, if a term occurs frequently in many documents regardless of their type, this weakens the uniqueness of the term from an information-theoretic point of view. This is measured by IDF as in Eq.~\ref{eq:idf}. 

\begin{equation}
    idf(t, D) = log \frac {|D|+1}{df(t) + 1}
    \label{eq:idf}
\end{equation}

IDF depends on Document Frequency (DF), Eq.~\ref{eq:df}, which is the number of documents containing the term $t$. IDF can be interpreted as soft self-information of the term $t$ where the probability distributions of the terms are mere approximations. Note that in the extreme case where $t$ presents in every document, $t$ does not carry any meaningful information, i.e. $idf(t,D)=0$.

TF-IDF is simply the TF measure scaled by IDF as defined in Eq.~\ref{eq:tf-idf}. 

\begin{equation}
    \text{tf\_idf}(t, D) = \mathrm{\overline{F}} \cdot  idf(t, D)
    \label{eq:tf-idf}
\end{equation}

TF-IDF can be used to weigh the elements of the DTM matrix $F$. With enough training samples, it is expected to extract the strength of the relationship between the type and the document via the classification of DTM weighted by TF-IDF. 

Similarly, Bag of Characters (BoC) and Bag of Words (BoW) are DTM matrices comprised of multiplicities of non-normalised terms, $f_{t,d}$, where the term $t$ is a \textit{character} for BoC and a \textit{word} for BoW.

The rows of DTM are feature vectors for each document. These are used to classify the documents. As long as the vector size is within a reasonable length, classical discriminative ML methods such as SVM or XGboost can be employed to find the boundaries between the class representations. 

Each SQL sentence is considered as the document, $d$. The definition of a term, $t$, affects the representation quality of DTM features extracted from $d$. A term might be defined as a single character in which case it is called monogram analysis. Monogram analysis of a text is the investigation of the statistical distribution of the characters at the finest level. It disregards the connection between the terms. There are also other levels of analysis which give more information about the relation of the characters such as bigram term analysis. 

If a sample SQLi payload is given as $d'$:  
\begin{lstlisting}[columns=fullflexible, language=SQL, breaklines]
SELECT * FROM users WHERE name = 'admin'--' AND password = ''
\end{lstlisting}
A monogram, bigram and trigram $t\in T^n$ would be in $T^1=\{\text{`S',`E',`L',...}\}$, $T^2=\{\text{`SE',`EL',`LE',...}\}$, $T^3=\{\text{`SEL',`ELE',`LEC',...}\}$, respectively.

\subsection{Transformer based models}
\label{sec:sub:transformers}
Transformers are a type of neural network architecture that have been highly successful in NLP tasks. They were first introduced in 2017 by Google \cite{vaswani2017attention} and are based on the idea of self-attention, where each word in a sentence attends to all the other words in the sentence to build a context representation. This allows Transformers to capture long-range dependencies and relationships between words more effectively than previous NLP models.

BERT (Bidirectional Encoder Representations from Transformers) \cite{devlin2018bert, turc2019} is a pre-trained language model for NLP tasks that is based on the encoder only Transformer architecture. BERT is bidirectional, meaning it can look at the context of a word in both directions (before and after the word) to better understand its meaning. It is pre-trained on a large corpus of text using two unsupervised learning tasks: masked language modelling and next-sentence prediction. Masked language modelling involves masking some of the words in a sentence and predicting them based on the surrounding context. Next sentence prediction involves predicting whether two sentences are consecutive in a text corpus or not. Once trained, BERT can be fine-tuned for specific NLP tasks such as sentiment analysis, question answering, and named entity recognition. Fine-tuning involves adding a task-specific layer on top of BERT and training it on a smaller labelled dataset. BERT has achieved state-of-the-art performance on several NLP benchmarks and has become a popular choice for various NLP applications.

Later in 2019, smaller BERT models and other variants were introduced \cite{turc2019, githubSmallBerts}. In this paper, we have adapted several variations of BERT models to SQLi detection tasks and investigated their performances in various aspects. One of the highest perfoming BERT model is incorporated into the proposed method.

\section{Proposed Method}
\label{sec:proposed}

\subsection{F1 Efficiency Metric}
\label{sec:sub:f1-efficiency}
As it will be demonstrated in the experiments, some of the methods have similar detection accuracies but have significantly different computation requirements during the inference. This creates a need for an objective measurement which takes both detection accuracy and  inference speed into consideration. This provides a practical middle ground between accuracy and speed. Another reason for this metric is, it could allow the system administrators who have limited computation power to spend on detection tasks to make an optimal decision on the SQLi detection method based on their computational restrictions. Moreover, in the future, this metric also can be used for an adaptive model selection method on dynamically changing computational resources. For those reasons, we introduce a new metric, \emph{F1 efficiency} (FE) as the weighted average of F1 and normalized inference time in ms ($l$) (Eq.~\ref{eq:f1-fe}):

\begin{equation}
    FE = \alpha F1 + (1-\alpha)l
    \label{eq:f1-fe}
\end{equation}
where $\alpha\in[0,1]$ and $l\in(0,1]$ is normalised inference time.

The \emph{FE} metric will be used in Section~\ref{sec:experiments} and Section~\ref{sec:sub:effective} to measure the effectiveness of the methods and design a system with optimum efficiency for limited computational resources.

\subsection{Feature extraction and Single NLP approach}
\label{sec:sub:feature-extraction}
The rows of the DTM matrix are used as feature vectors corresponding to each SQL statement and the columns are the terms, $t$. A DTM might be constructed in different ways. To find the optimal feature set we have tested various DTM matrices. As explained in Section~\ref{sec:NLPDetectionOfSQLInjectionAttack}, when the raw count of the characters is used, it is called BoC. In this case, DTM of BoC, $\mathrm{F}^{BoC}$, can be defined more explicitly as:
\begin{equation}
    \mathrm{F}^{BoC} = F , ~t\in T^{1}
    \label{eq:dtm-boc}
\end{equation}
Similarly, the DTM of BoW, $\mathrm{F}^{BoW}$, can be obtained by tokenizing each word as such $t\in T^W$ where $T^W=\{\text{`SELECT'},\text{`FROM'},\text{`users'},...\}$.

The other features are $\overline{\mathrm{F}}$ which is normalised raw counts as in Eq.\ref{eq:tf}. TF-IDF matrix $\mathrm{\widehat{F}^{n}}$ can be defined as: 
\begin{equation}
    \mathrm{\widehat{F}^{n}} = [\mathrm{tf\_idf}(j,i)], ~t\in T^n
    \label{eq:F-tfidf}
\end{equation}
These features, i.e., 
\begin{equation}
    \{\mathrm{F},~\overline{\mathrm{F}},~\mathrm{F}^{BoC},~\mathrm{F}^{BoW},~\mathrm{\widehat{F}^{n}}\}
    \label{eq:features-list}
\end{equation}
are tested in various experimental settings.

\begin{figure}
	\centering
	\includegraphics[scale=0.3]{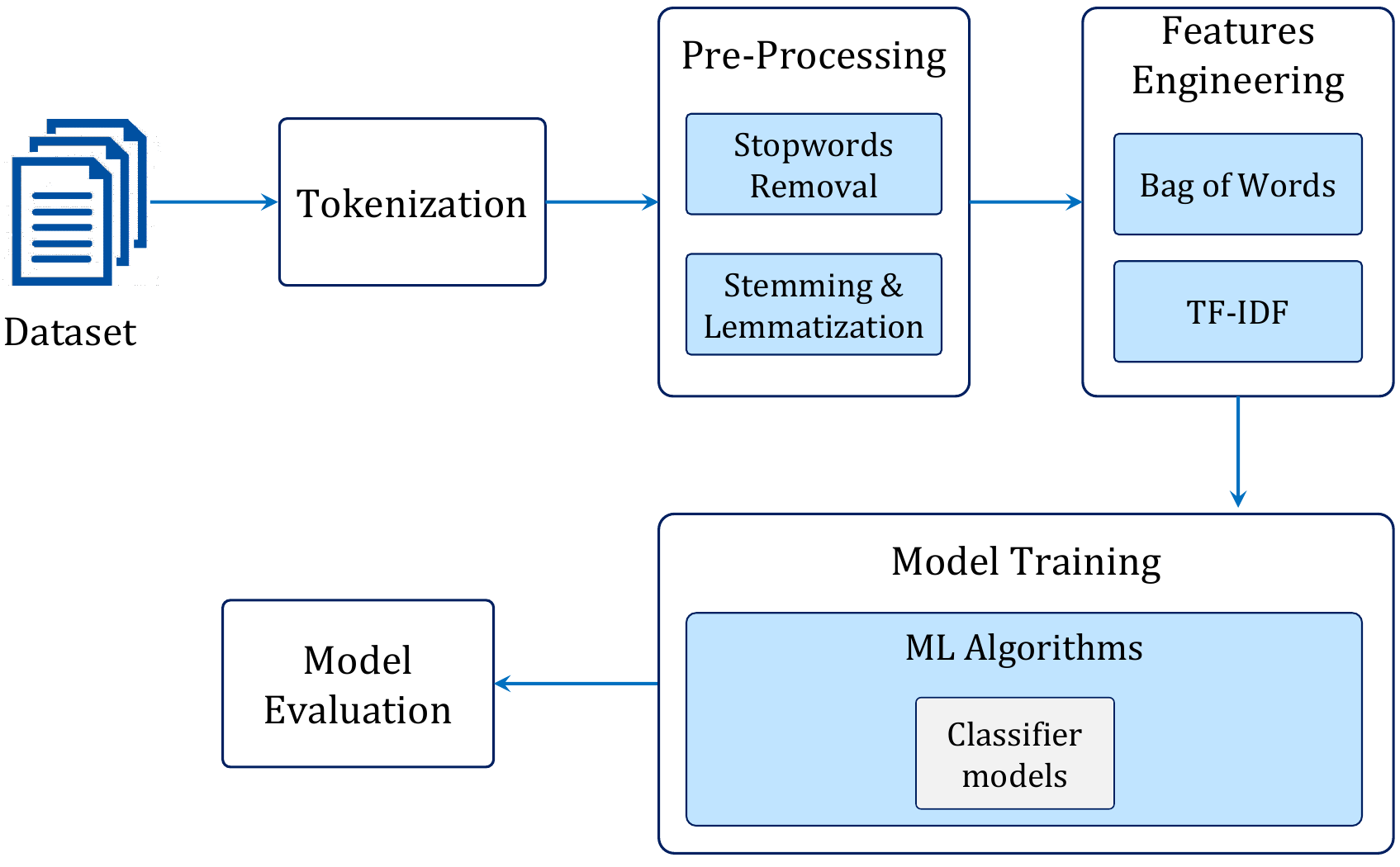}
	\caption{The single NLP process which is combining classical ML classifiers with various features given in Eq.\ref{eq:features-list}. (TF-IDF based and Bag of char/word based features) The SQL payload is first tokenized and cleaned in the pre-processing phase. In the next stage, features are extracted with the selected method. Then the features are used in the selected model's training.}
	\label{fig:single_nlp_process}
\end{figure}

Various combinations of classical ML classifiers and features in Eq.\ref{eq:features-list} are investigated as the ``Single NLP" approach. The general flow of the Single NLP approach is given in Fig.~\ref{fig:single_nlp_process}. The single NLP approach comprises of classical ML classifier and
the features extracted by the selected method. Their detection accuracy (F1 score) and inference speed per sample are demonstrated as purple bars in Fig.~\ref{fig:all-F1-vs-inference}.

Overall, these methods have the lowest inference latencies but slightly lower detection accuracy.

\subsection{Ensemble Strategies}
\label{sec:sub:ensemble-strategies}
In order to find the optimal combination of features and classification algorithms, four different ensemble settings are tested.

\begin{figure}
	\centering
	\includegraphics[width=0.5\textwidth]{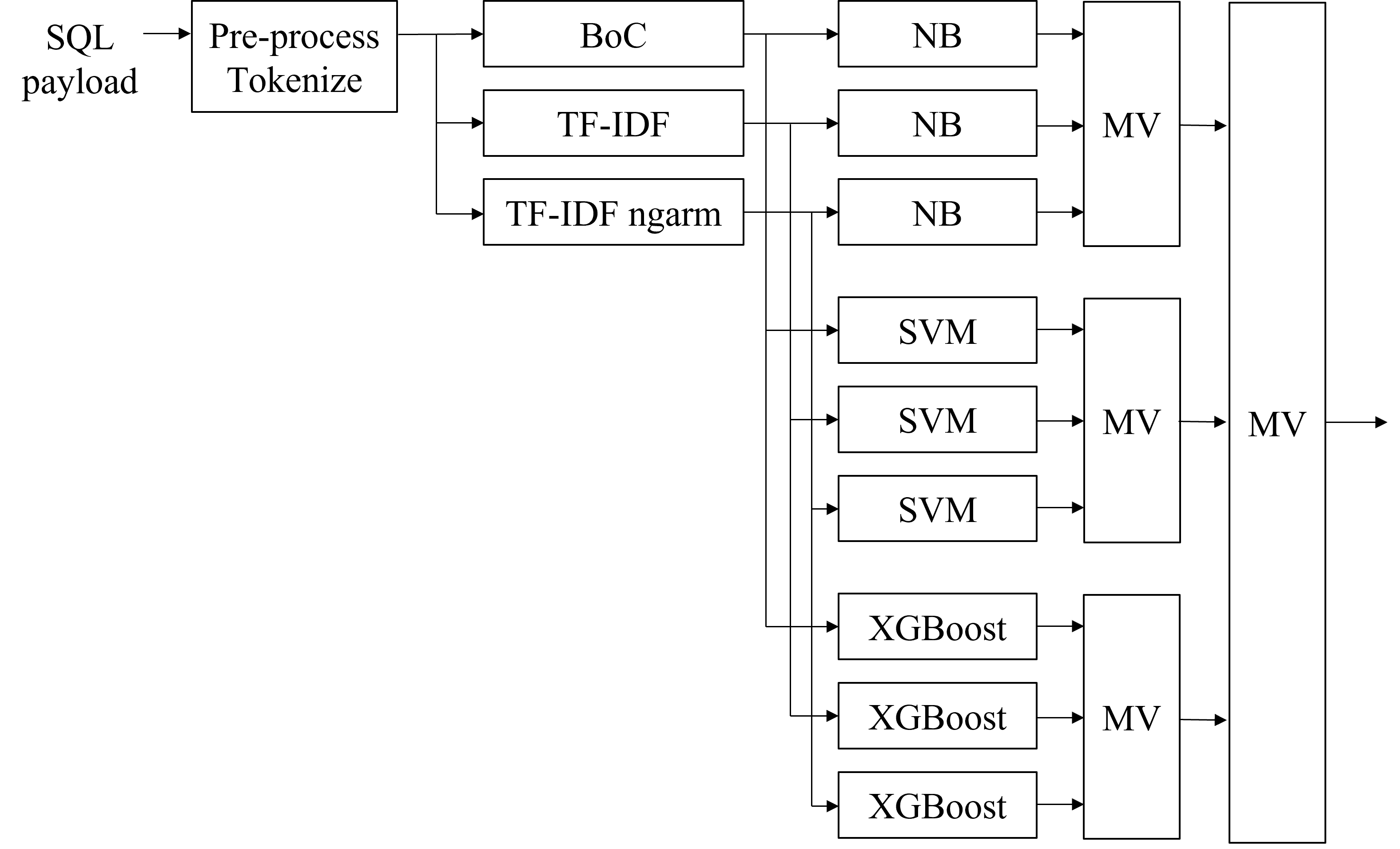}
	\caption{Proposed ensemble approach 1. Each classifier type is used with multiple features. }
	\label{fig:ensemble_1}
\end{figure}

In the first ensemble setting as depicted in Fig.~\ref{fig:ensemble_1}, the features given in Eq.\ref{eq:features-list} are extracted and individually fed to classification tools (i.e., \naive Bayes, Xgboost and SVM). The Majority Voting (MV) of the classification results of each classification method is taken as the final decision. In this strategy, the performance of the decisions is based on that of each classification method.

\begin{figure}
	\centering
	\includegraphics[width=0.5\textwidth]{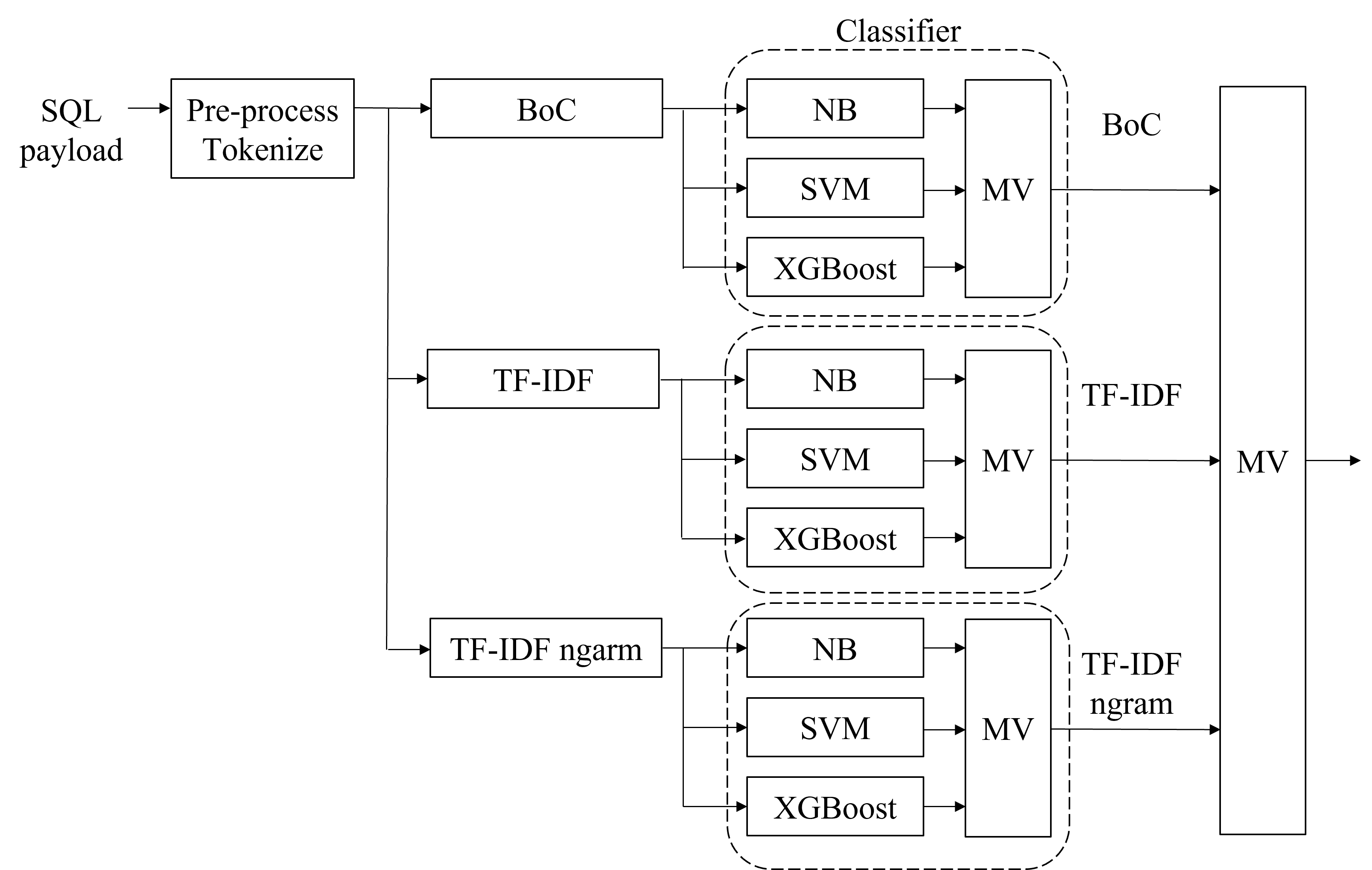}
	\caption{Proposed ensemble approach 2. Each feature set type is classified with multiple classifiers.}
	\label{fig:ensemble_2}
\end{figure}

In the second ensemble setting (Fig.~\ref{fig:ensemble_2}), unlike the previous one, the decisions of each ML method are majority-voted. Therefore the performance of each of the three decisions is mostly based on the features.

\begin{figure}
	\centering
	\includegraphics[width=0.5\textwidth]{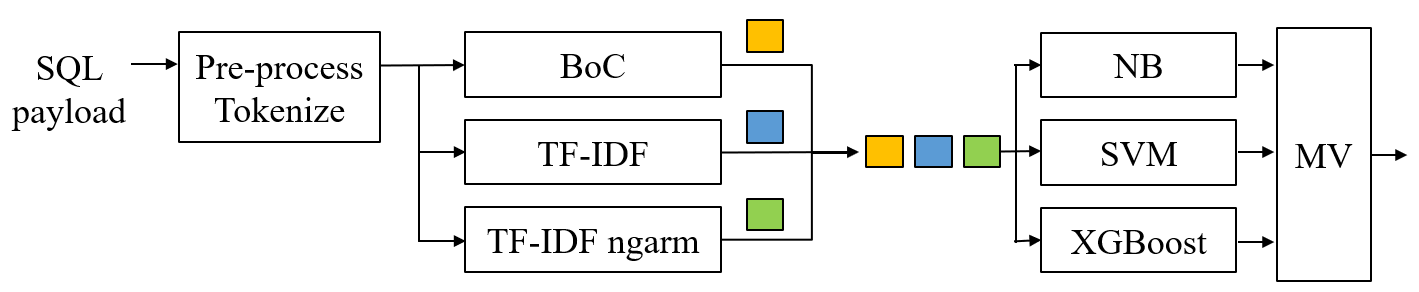}
	\caption{Proposed ensemble approach 3. This is similar to Ensemble 1 approach but in this case, the features are concatenated before being classified.}
	\label{fig:ensemble_3}
\end{figure}

The final setting, "ensemble 3" (Fig.~\ref{fig:ensemble_3} fuses the feature set into a single feature vector. In this way, each ML method can learn the most meaningful combination of the feature elements.

\subsection{The Proposed cascade SQLi detector}
\label{sec:sub:two-stage}

The utilization of multi-staged classification approaches in previous studies, such as \cite{viola2001rapid, two-stage-text, two-stage-intrusion}, has demonstrated the efficacy of combining fast and accurate models. Building upon this approach, we propose a two-stage cascade SQLi detection system that achieves high-accuracy detection of SQLi at high speeds.

As highlighted in Section~\ref{sec:experiments}, classical machine learning classifiers generally exhibit lower computational costs compared to the latest transformer-based NLP methods. However, transformers exhibit higher classification accuracy when compared to classical NLP methods. To capitalize on the speed of classical methods and the accuracy of transformers, we introduce a two-stage cascade SQLi detection system, as illustrated in Fig.~\ref{fig:two_stage}.

In the first stage, a fast classification method is employed to swiftly eliminate the majority of negative samples with minimal computation. The second stage then re-examines the positively labeled samples from the first stage to reduce false positives. The overall system is estimated to possess an average speed akin to classical ML methods while maintaining comparable detection accuracy to transformers, as shown in Eq.~\ref{eq:approx}.

The first stage classifier (Passive Aggressive Classifier\cite{crammer2006online}) is trained to be more sensitive to positive samples, aiming to minimize false negatives while allowing for higher false positives. To achieve this, we scaled the weight of positive samples by a factor of 1000 during the training of the first stage model. During the initial inference, a threshold of -0.3 was used to capture even slightly suspicious SQL payloads. These fine-tuning techniques led to a reduction in false negatives from around $\sim10$ to $\sim6$, with an increase in false positives to $\sim60$, which are further analysed by the transformer-based model in the second stage.

It is important to note that the inference latency of the proposed method is dependent on the ratio of attacks. As such, the estimation of inference latency is carried out for a realistic scenario, as detailed in Section~\ref{sec:sub:effective}.

\section{Experiments and Performance Analysis}
\label{sec:experiments}

The experimental setup is designed to reveal the comparative performance of the proposed  and other baseline methods in terms of classification accuracy, training and inference speed. As shown in Fig.~\ref{fig:single_nlp_process}, first, the dataset is tokenized and preprocessed. The features are extracted from the preprocessed data. The baseline ML algorithms and the proposed ensemble methods are trained and tested on the features coming from the previous step.

For the transformer-based models, preprocessing is performed with the selected model's respective preprocessing method.

\begin{figure*}[ht]
     \centering
     \begin{subfigure}{0.49\textwidth}
         \centering
         \includegraphics[width=\textwidth]{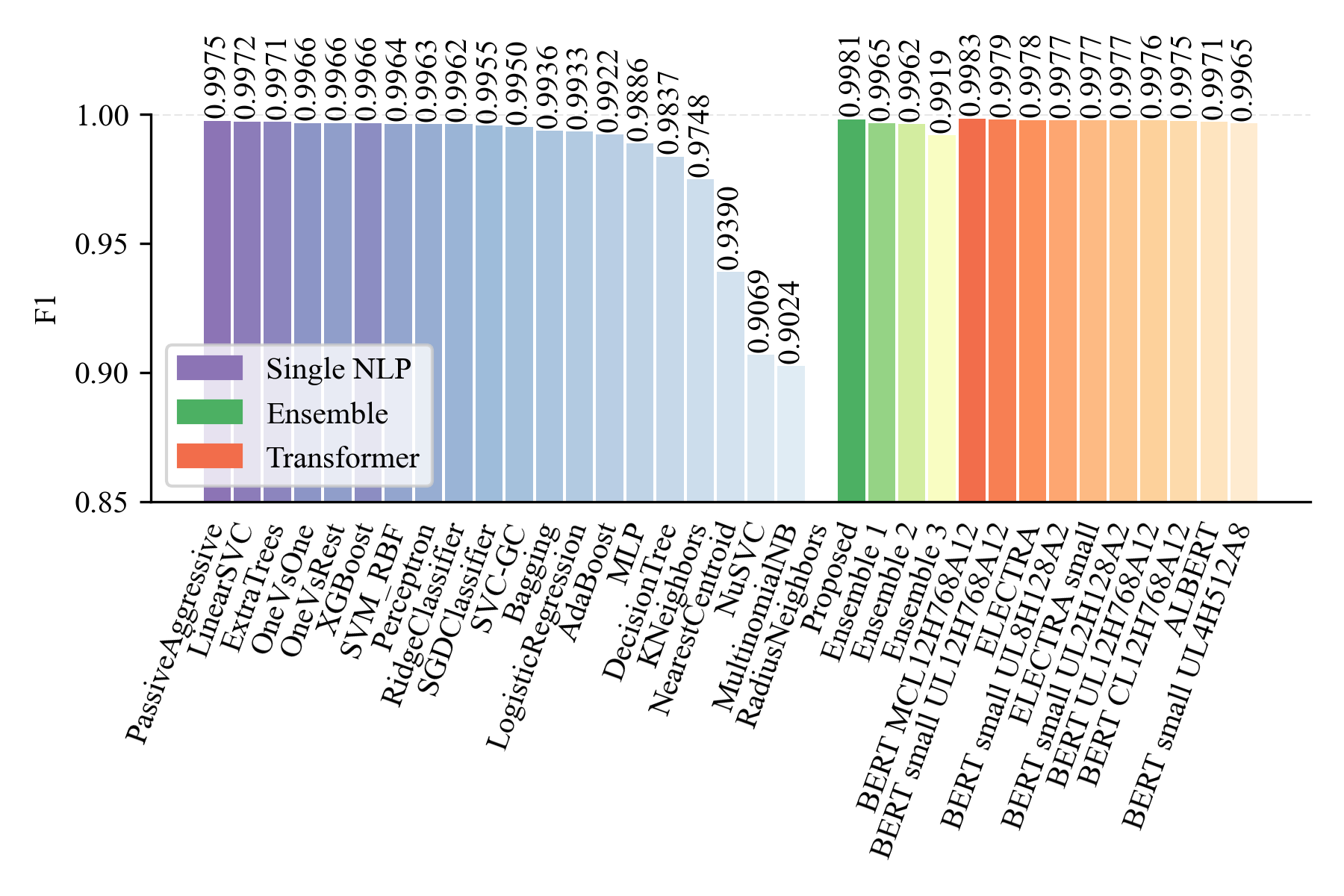}
         \caption{}
         \label{fig:sub:F1-all}
     \end{subfigure}
     \hfill
     \begin{subfigure}{0.49\textwidth}
         \centering
         \includegraphics[width=\textwidth]{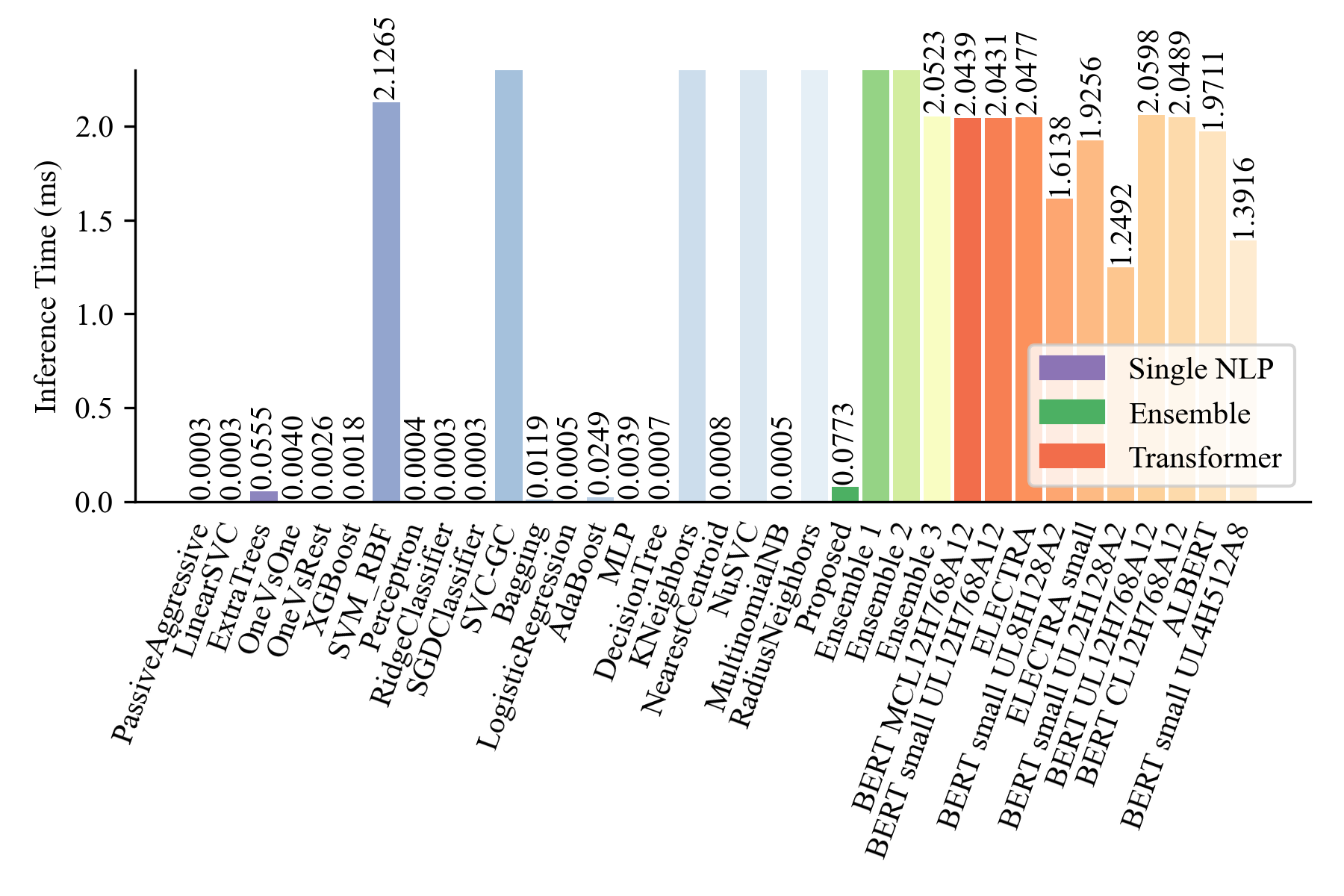}
         \caption{}
         \label{fig:sub:inference-all}
     \end{subfigure}
        \caption{All methods are compared with respect to (a) F1 and (b) inference latency scores. Single NLP, Ensemble and Transformer methods are shown as shades of purple, green and orange. The methods are sorted by F1 scores within each method category. Each method uses the same colour throughout the paper. The F1 scores of most of the methods are well over 95\%. However, their inference latencies are significantly different. Overall, the transformer-based models give the highest detection accuracies whereas the classical ML-based NLP methods have the highest inference speeds.}
        \label{fig:all-F1-vs-inference}
\end{figure*}

\subsection{Test environment}
\label{sec:sub:test-environment}
The experiments were conducted using Google Colab, leveraging a high-end GPU (A100 SXM 40GB) for computational support. We recorded the training and inference times for both the Large Language Models (LLMs) and the classical models. To ensure robustness and reduce the potential influence of random factors, each experiment was replicated ten times using reshuffled datasets. The presented results represent the average of these repeated tests.

It is important to note that for the BERT model, inferences were conducted in batches. To calculate the average inference time per sample, the total inference time was divided by the batch size. However, when inferring on a single sample, one may observe a slower inference latency than that derived from the batch-based calculation.

A publicly available SQLi detection dataset with 30.609 samples is used \cite{kaggleDS}. The dataset has 11.341 positive, 19.268 negative samples.

\subsection{Comparison of Overall detection performances}
\label{sec:sub:overall-performances}
The detection accuracy (F1 score) and the average inference latency per sample of each method are evaluated (See Fig.\ref{fig:all-F1-vs-inference}). The figure shows that each of the three model groups (Single NLP, Ensemble and Transformer) has members with high F1 scores ($>$99\%). However, there is a notable difference in their inference latencies (0.0003ms to 2ms). These are elaborated in the following sections. Each model group is investigated in a separate section.

\subsection{Single NLP Algorithms}
\label{sec:sub:single-NLP}
The first part of the experiments is performed with the simplest form of detection approaches. A single ML tool is used with a single feature vector at a time. Due to their simplicity in terms of computational complexity relative to the others, this type of approach is expected to be generally less accurate but also computationally less demanding. 
The features listed in Eq.~\ref{eq:features-list} are extracted using BoC, BoW, TF-IDF and TF-IDF n-gram methods. These are used in the classification of SQLi statements. The classification methods used in our study encompass various models including XGBoost, MultinomialNB, Support Vector Machines with Radial Basis Function (SVM-RBF), Multilayer Perceptron (MLPClassifier), K-Nearest Neighbors (KNeighborsClassifier), Nearest Centroid (NearestCentroid), Radius Neighbors Classifier (RadiusNeighborsClassifier), Support Vector Classification with Gram matrix (SVC-GC), Nu-Support Vector Classification (NuSVC), Linear Support Vector Classification (LinearSVC), Decision Tree Classifier (DecisionTreeClassifier), AdaBoost Classifier (AdaBoostClassifier), Bagging Classifier (BaggingClassifier), Extra Trees Classifier (ExtraTreesClassifier), Ridge Classifier (RidgeClassifier), Stochastic Gradient Descent Classifier (SGDClassifier), Perceptron, Logistic Regression (LogisticRegression), Passive Aggressive Classifier (PassiveAggressiveClassifier), One-vs-Rest Classifier (OneVsRestClassifier), and One-vs-One Classifier (OneVsOneClassifier). 

The results are given in Table~\ref{tab:single} with the best performance scores highlighted for each category. The table demonstrates that the PassiveAggressive classifier achieved the highest accuracy of 0.9981 and the best F1 score of 0.9975. Perceptron, LinearSVC and ExtraTrees classifiers demonstrated a strong performance with the highest True Positive (TP) and True Negative (TN) values respectively.

RadiusNeighbors and KNeighbors classifiers had considerably slower inference times, over 31 milliseconds each, making them potentially less suitable for time-sensitive applications.

\begin{figure*}[!th]
\centering
\includegraphics[width=0.98\textwidth]{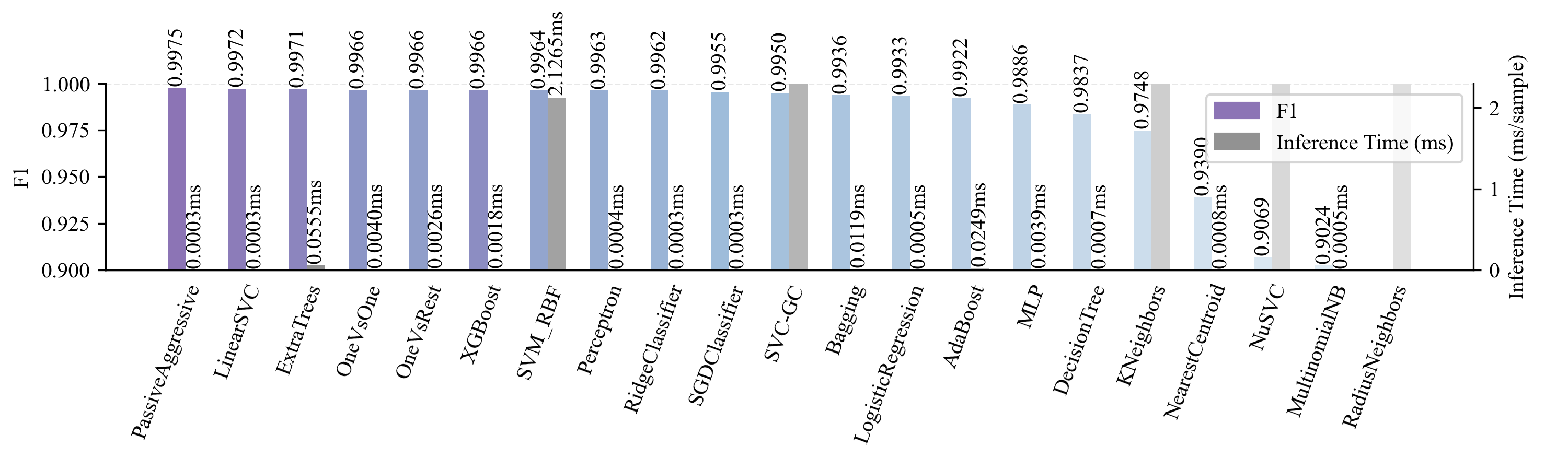}
\caption{Performance comparison of machine learning models depicted in a dual-bar plot: a) F1 scores represented by purple bars, and b) inference latencies denoted by grayscale bars. The initial two methods, PAC (Passive Aggressive Classifier) and LinearSVC (Linear Support Vector Classification), exhibit comparable performance in terms of both F1 scores and inference latency.}
\label{fig:single_f1_time}
\end{figure*}

For a visual comparison, F1 scores and inference times are plotted in Fig.~\ref{fig:single_f1_time}. The results show a trade-off between F1 and speed across different classifiers. The figure shows that the closest competitor of the PassiveAggressive classifier in terms of F1 score and inference time is LinearSVC. Although the PassiveAggressive classifier has the highest F1 score, LinearSVC manages to keep up with a slightly lower F1 score, while having the lowest inference time among all the classifiers. 

As a result of these findings, the first stage model of the proposed system is selected as the PassiveAggressive classifier, since it has both low inference latency and high detection accuracy.

\begin{table*}[htbp]
\centering
\caption{Performance metrics of the machine learning models are presented, based on the average of 10 independent test runs. The Passive Aggressive Classifier (PAC) emerges as the most suitable model in terms of inference speed and detection accuracy.}
\label{tab:single}
\begin{tabular}{lrrrrrrrrrr}
\toprule
{Method} & {Accuracy} & {Precision} & {Recall} & {F1} & {TP} & {TN} & {FP} & {FN} & \makecell{Training\\Time (ms)} & \makecell{Inference\\Time (ms)} \\
\midrule
AdaBoost & 0.9942 & 0.9962 & 0.9882 & 0.9922 & 2240 & 3846 & 9 & 27 & 0.8238 & 0.0249 \\
Bagging & 0.9953 & 0.9969 & 0.9904 & 0.9936 & 2245 & 3848 & 7 & 22 & 2.3887 & 0.0119 \\
DecisionTree & 0.9880 & 0.9950 & 0.9726 & 0.9837 & 2205 & 3844 & 11 & 62 & 0.0727 & 0.0007 \\
ExtraTrees & 0.9978 & 0.9995 & 0.9947 & 0.9971 & 2255 & 3854 & 1 & 12 & 0.4283 & 0.0555 \\
KNeighbors & 0.9815 & 0.9829 & 0.9668 & 0.9748 & 2192 & 3817 & 38 & 75 & 0.0005 & 31.6636 \\
LinearSVC & 0.9979 & 0.9996 & 0.9948 & 0.9972 & 2255 & 3854 & 1 & 12 & 0.0052 & \textbf{0.0003} \\
LogisticRegression & 0.9951 & 0.9986 & 0.9881 & 0.9933 & 2240 & 3852 & 3 & 27 & 0.0684 & 0.0005 \\
MLP & 0.9917 & 0.9982 & 0.9793 & 0.9886 & 2220 & 3851 & 4 & 47 & 1.8507 & 0.0039 \\
MultinomialNB & 0.9209 & 0.8309 & 0.9876 & 0.9024 & 2239 & 3399 & 456 & 28 & 0.0007 & 0.0005 \\
NearestCentroid & 0.9570 & 0.9882 & 0.8945 & 0.9390 & 2028 & 3831 & 24 & 239 & 0.0011 & 0.0008 \\
NuSVC & 0.9369 & \textbf{0.9996} & 0.8300 & 0.9069 & 1882 & \textbf{3854} & \textbf{1} & 385 & 15.5939 & 7.8269 \\
OneVsOne & 0.9975 & 0.9992 & 0.9940 & 0.9966 & 2253 & 3853 & 2 & 14 & 0.3494 & 0.0040 \\
OneVsRest & 0.9975 & 0.9992 & 0.9940 & 0.9966 & 2253 & 3853 & 2 & 14 & 0.3745 & 0.0026 \\
PassiveAggressive & \textbf{0.9981} & 0.9994 & 0.9956 & \textbf{0.9975} & 2257 & 3854 & 1 & 10 & 0.0057 & 0.0003 \\
Perceptron & 0.9972 & 0.9965 & \textbf{0.9960} & 0.9963 & \textbf{2258} & 3847 & 8 & \textbf{9} & 0.0026 & 0.0004 \\
RadiusNeighbors & 0.8266 & 0.6894 & 0.9682 & 0.8053 & 2195 & 2866 & 989 & 72 & \textbf{0.0004} & 31.6458 \\
RidgeClassifier & 0.9972 & 0.9994 & 0.9930 & 0.9962 & 2251 & 3854 & 1 & 16 & 0.0690 & 0.0003 \\
SGDClassifier & 0.9967 & 0.9994 & 0.9917 & 0.9955 & 2248 & 3854 & 1 & 19 & 0.0051 & 0.0003 \\
SVC-GC & 0.9963 & 0.9993 & 0.9907 & 0.9950 & 2246 & 3853 & 2 & 21 & 11.1949 & 3.8048 \\
SVM\_RBF & 0.9973 & 0.9994 & 0.9934 & 0.9964 & 2252 & 3854 & 1 & 15 & 5.0202 & 2.1265 \\
XGBoost & 0.9975 & 0.9992 & 0.9940 & 0.9966 & 2253 & 3853 & 2 & 14 & 0.4057 & 0.0018 \\
\bottomrule
\end{tabular}
\end{table*}

\subsection{Ensemble NLP Algorithms}
\label{sec:sub:ensemble_nlp_results}
\begin{figure}[!th]
\centering
\includegraphics[width=0.5\textwidth]{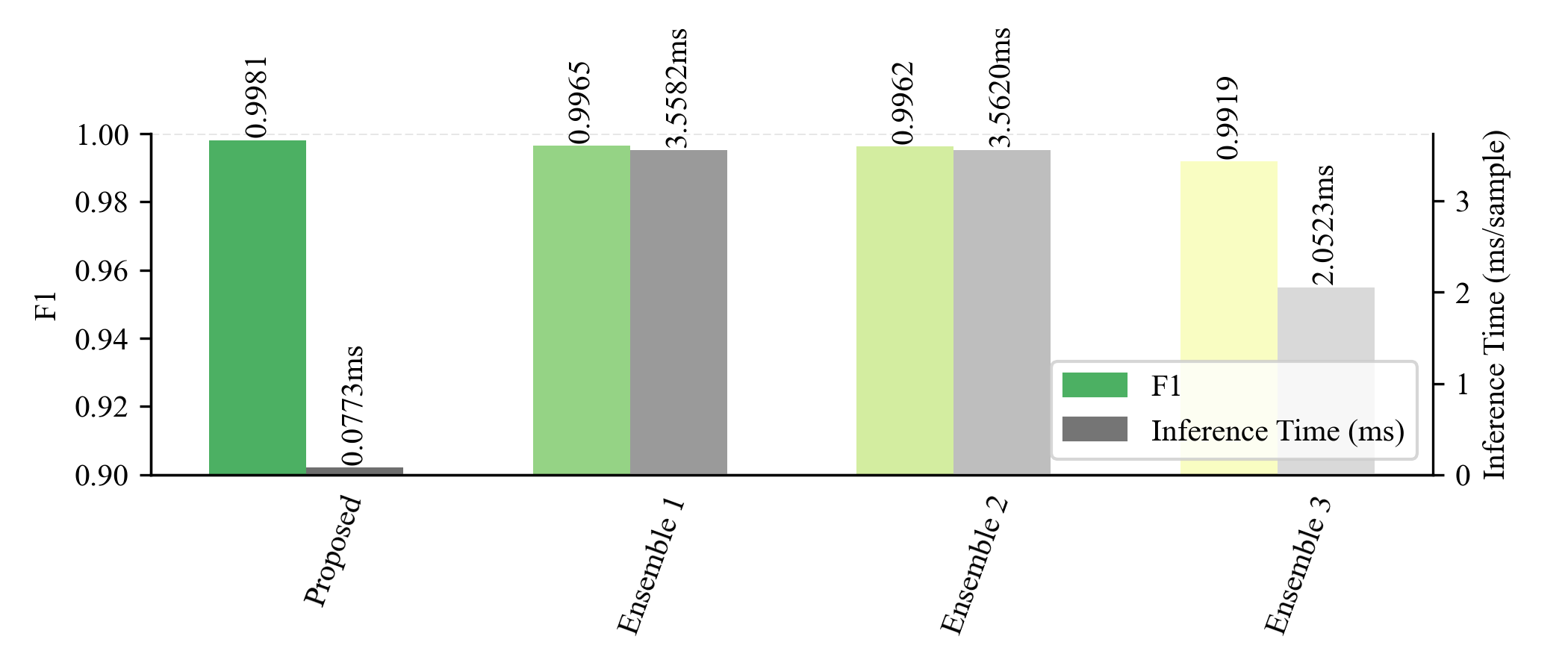}
\caption{Performance comparison of ensemble models depicted in a dual-bar plot: a) F1 scores represented by purple bars, and b) inference latencies denoted by grayscale bars. Among all ensembles, the proposed two-stage method stands out with its higher detection accuracy and marginally lower inference latency.}
\label{fig:ensemble_f1_time}
\end{figure}

Section~\ref{sec:sub:ensemble-strategies} presents four distinct ensemble approaches, with Fig.~\ref{fig:ensemble_f1_time} illustrating the ensemble methods ranked by F1 scores. As evident from the plot, the proposed cascade SQLi detection method achieves the highest F1 score, while also exhibiting significantly lower computation costs.

The proposed method is rigorously evaluated through six rounds of testing, wherein the dataset is reshuffled at each iteration. The inference latency of the proposed method is estimated as elaborated in Section~\ref{sec:sub:effective} for each of the six test rounds, and the results are averaged thereafter.

Table~\ref{tab:ensemble} showcases that the Ensemble-1 method yields the highest F1 score following the proposed cascade SQLi detection method. However, the inference times of the ensemble model are marginally higher than those of the BERT models, rendering them less practical in terms of preference.

\begin{table*}[htbp]
\centering
\caption{Performance metrics of various ensemble NLP models, averaged over 10 independent test runs, are presented. The proposed model demonstrates superior performance in comparison to the other ensemble models.}
\label{tab:ensemble}
\begin{tabular}{lrrrrrrrrrr}
\toprule
{Method} & {Accuracy} & {Precision} & {Recall} & {F1} & {TP} & {TN} & {FP} & {FN} &  \makecell{Training\\Time (ms)} & \makecell{Inference\\Time (ms)} \\
\midrule
Ensemble 1 & 0.9974 & \textbf{0.9996} & 0.9934 & 0.9965 & 2252 & \textbf{3854} & \textbf{1} & 15 & 22.5116 & 3.5582 \\
Ensemble 2 & 0.9972 & 0.9996 & 0.9929 & 0.9962 & 2251 & \textbf{3854} & \textbf{1} & 16 & 22.5116 & 3.5620 \\
Ensemble 3 & 0.9940 & 0.9942 & 0.9896 & 0.9919 & 2243 & 3842 & 13 & 24 & \textbf{4.9306} & 2.0523 \\
Proposed & \textbf{0.9986} & 0.9996 & \textbf{0.9966} & \textbf{0.9981} & \textbf{2259} & \textbf{3854} & \textbf{1} & \textbf{8} & 68.7000 & \textbf{0.0773} \\
\bottomrule
\end{tabular}
\end{table*}

In our subsequent analysis, we delve into the trade-off between F1 score and inference latency through the lens of the FE measure. When we calculate the FE metric (as per Eq.\ref{eq:f1-fe}) for the ensemble methods with varying values of $\alpha$ weight, we observe a reshuffling of the methods' rankings, as depicted in Fig~\ref{fig:FE-all}. The results reveal that at an $\alpha$ weight of 0.98, both transformer-based and ensemble models slip down the list due to their higher computational requirements. In such a scenario, our proposed method becomes a more advantageous choice.

\subsection{Transformer-Based NLP Algorithms}
\label{sec:sub:transformer_results}
We tested one of the most prominent transformer-based NLP architectures, BERT \cite{devlin2018bert, turc2019}, and its variants against the SQLi detection task. 
The experimental results are presented in Table~\ref{tab:transformer}. As anticipated, these models outperform classical ML methods in terms of detection accuracy. Notably, the Electra base model exhibits the highest recall rate on the dataset, indicating the fewest missed detections among all models. Because of its superior performance in minimizing missed detections, it has been selected as the second stage model in our proposed method. 

\begin{figure}[!thb]
\centering
\includegraphics[width=0.5\textwidth]{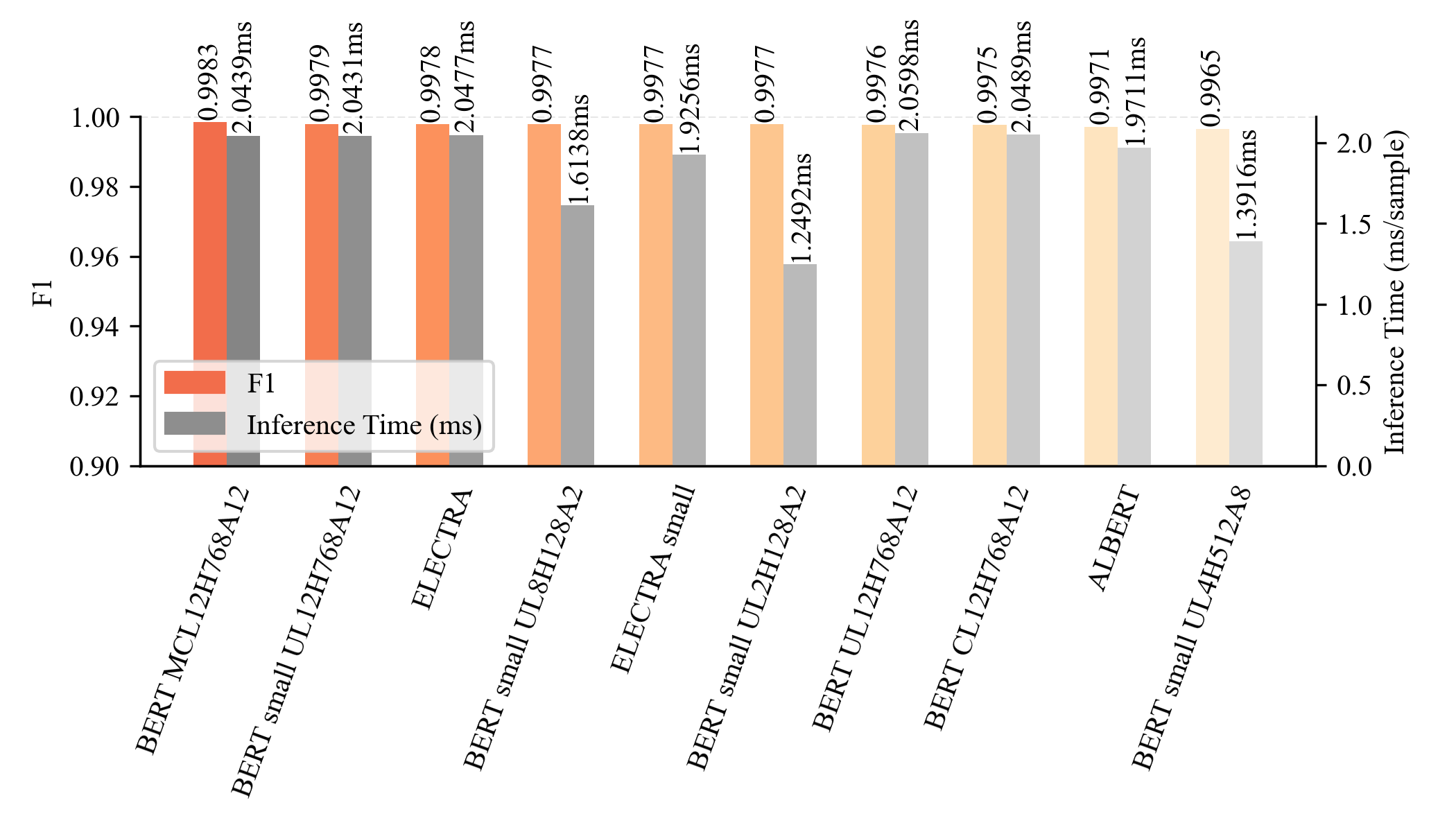}
\caption{This figure displays the classification and inference time performance of transformer-based NLP methods. a) F1 scores are depicted with purple bars and b) inference latencies are represented with grayscale bars. As the results illustrate, ELECTRA accomplishes the highest recall, while BERT, with model parameters L=12, H=768, A=12, presents the top performance in terms of accuracy and F1 scores.}
\label{fig:transformer_f1_time}
\end{figure}

The table further highlights that the BERT multi-cased model, configured with hyperparameters L=12, H=768, and A=12, achieves the highest accuracy and F1 scores. However, these high-performing models come with a significant downside—impractically high computational demands, as illustrated in Fig.~\ref{fig:transformer_f1_time}. Each model requires approximately 2ms to analyze each SQL statement, a timeframe significantly longer than the mere microseconds required by classical ML methods.

When the F1 score is assigned 98\% importance weight ($\alpha=0.98$), the ranking of the methods changes, as shown in Fig.~\ref{fig:FE-all}. Under these conditions, the method with the highest FE score—multi-cased BERT with L=12, H=768, A=12—scores lower than several classical ML methods.

\begin{table*}[htbp]
\centering
\caption{Table presents the performance metrics of various transformer-based NLP models, averaged over 10 independent test runs. The results demonstrate that ELECTRA achieves the highest recall, while BERT with model parameters L=12, H=768, A=12 exhibits the highest accuracy and F1 scores.}
\label{tab:transformer}
\begin{tabular}{lrrrrrrrrrr}
\toprule
{Method} & {Accuracy} & {Precision} & {Recall} & {F1} & {TP} & {TN} & {FP} & {FN} & \makecell{Training\\Time (ms)} & \makecell{Inference\\Time (ms)} \\
\midrule
ALBERT & 0.9978 & 0.9983 & 0.9958 & 0.9971 & 2258 & 3851 & 4 & 10 & 12.3332 & 1.9711 \\
BERT CL12H768A12 & 0.9982 & 0.9979 & 0.9972 & 0.9975 & 2261 & 3850 & 5 & 6 & 12.8636 & 2.0489 \\
BERT UL12H768A12 & 0.9983 & 0.9979 & 0.9974 & 0.9976 & 2262 & 3850 & 5 & 6 & 13.0167 & 2.0598 \\
BERT MCL12H768A12 & \textbf{0.9988} & 0.9993 & 0.9974 & \textbf{0.9983} & 2262 & 3853 & 2 & 6 & 14.5866 & 2.0439 \\
ELECTRA & 0.9984 & 0.9979 & \textbf{0.9978} & 0.9978 & \textbf{2262} & 3850 & 5 & \textbf{5} & 12.8706 & 2.0477 \\
ELECTRA small & 0.9983 & 0.9986 & 0.9968 & 0.9977 & 2260 & 3851 & 3 & 7 & 10.7674 & 1.9256 \\
BERT small UL12H768A12 & 0.9984 & 0.9983 & 0.9974 & 0.9979 & 2262 & 3851 & 4 & 6 & 12.7917 & 2.0431 \\
BERT small UL2H128A2 & 0.9983 & \textbf{0.9994} & 0.9960 & 0.9977 & 2258 & \textbf{3853} & \textbf{1} & 9 & \textbf{4.7080} & \textbf{1.2492} \\
BERT small UL4H512A8 & 0.9974 & 0.9960 & 0.9970 & 0.9965 & 2261 & 3845 & 9 & 7 & 6.1357 & 1.3916 \\
BERT small UL8H128A2 & 0.9983 & 0.9987 & 0.9968 & 0.9977 & 2260 & 3852 & 3 & 7 & 8.2361 & 1.6138 \\
\bottomrule
\end{tabular}
\end{table*}

The bar plots illustrated in Fig.\ref{fig:sub:FE-100-all} and Fig.\ref{fig:sub:FE-90-all} convey that when $\alpha\leq0.98$, the SVM-based methods, ensemble approaches, and transformer models lag behind most of the other methods due to their substantial computational demands. In this scenario, the Passive Aggressive classifier emerges as the most suitable model for the first stage of classification.

\subsection{Effective Computing Requirements}
\label{sec:sub:effective}
In the previous sections, the average inference latencies for each method are given. In this section, we calculate the effective compute time requirement of the proposed cascade SQLi detection, two-stage method, introduced in Section~\ref{sec:sub:two-stage} and depicted in Fig.\ref{fig:two_stage}. 

The inference latency of the proposed system is contingent upon the attack ratio, necessitating the calculation of effective inference latency rather than experimental demonstration. In a realistic scenario, only 3.3\% of the SQL queries are identified as actual attacks, as reported by OWASP \cite{owasp2021}. This implies that the second stage of the proposed method will be triggered infrequently, resulting in a speed gain. Conversely, the inference latencies for other methods in the comparison need not be calculated, as they remain unchanged regardless of whether it is an attack or not.

The effective computing time depends on its parts. The first part is the lightweight detection method which is fast but rises some false positives occasionally. It eliminates most of the negative SQL statements. The purpose of the method in the second stage is to capture the false positives of the first part and minimise the overall error rate. The second stage method is only triggered by positive classification decisions, $D\in\{0,1\}$, coming from the first stage. The probability $p(D=1)$ of the first part producing a positive decision depends on its false detection probability $p(D=1|C=0)$ and the apriori probability of positive SQL attacks in all SQL statements $p(C=1)$ as in Eq.~\ref{eq:bayes}. In the equation, $C\in\{0,1\}$ stands for the true classes.

\begin{align}
    p(D=1) & =  \sum_{i\in\{0,1\}}{p(D=1|C=i)p(C=i)} \label{eq:bayes}\\
    & \approx \text{FPR}\times(1-p(C=1)) + \text{Recall}\times(p(C=1))
    \label{eq:approx}
\end{align}

We can estimate $p(D=1|C)$ statistically based on our test results as in Eq.\ref{eq:approx}. The false positive rate (fall-out = FP/(TN+FP) ) is an approximation of $p(D=1|C=0)$. Similarly, True positive rate (recall = TP/(TP+FN) ) approximates to $p(D=1|C=1)$. However, we don't have the prior probability of positive attacks, $p(C=1)$. We used OWASP TOP 10 report \cite{owasp2021} as an approximation of the prior. The report's section 'A3 Injection' states that the average incident rate of the injection attacks (SQL + XSS + External Control of File Name or Path) altogether is $3.3\%$. When we take $p(C=1)\approx 0.033$, $p(D=1)$ can be estimated with FPR and recall scores of the first stage method which are 0.0157 and 0.9973 respectively as in Eq.\ref{eq:approx_numeric}. The result found in Eq.\ref{eq:approx_result} is slightly higher than the $3.3\%$ average incident rate because of the lower recall score of the first stage method. 

\begin{align}
    p(D=1) & \approx 0.0157\times(1-0.033) + 0.9973\times0.033
    \label{eq:approx_numeric}\\
    & \approx 0.04811
    \label{eq:approx_result}
\end{align}
This means that the second stage method will be triggered approximately $4.811\%$ of the time. Relying on this information, we can calculate the effective speed of the proposed system as in Eq.\ref{eq:system_inference_eq}
\begin{align}
    \mathscr{T} & = \mathscr{T}_1\times (p(D=1)+p(D=0))  + \mathscr{T}_2\times p(D=1) 
    \label{eq:system_inference_eq}\\
    & \approx 0.0988 ms
    \label{eq:system_inference_numeric}
\end{align}
where $\mathscr{T}_n$ is the average inference time per SQL statement of the stage n method. In this case, they were measured as $\mathscr{T}_1=0.000314$ and  $\mathscr{T}_2=2.047714ms$. The average effective inference time of the whole 2-stage system (see Eq.\ref{eq:system_inference_numeric}) is estimated to be 0.0988 ms, which is $\mathbf{\approx20\times }$ faster than the transformer-based method with the best accuracy. Our experiments show that the estimated speed and accuracy of the proposed system are in line with the test results given in Table~\ref{tab:ensemble}.

\section{Conclusion}
\label{sec:Conclusion}
Throughout this paper, we have comprehensively explored and compared 35 different methods, ranging from traditional machine learning models to the most recent transformer-based models, all in the context of SQLi detection. Our analysis has shown that while transformer models may yield higher accuracy, they often come with the drawback of significantly higher computational demand. On the other hand, classical machine learning models, particularly the PassiveAggressive classifier, demonstrated remarkable efficiency, balancing high detection accuracy with lower inference latency.

To capitalize on the strengths of both categories of models, we introduced a novel cascaded SQLi detection model. This model integrates classical ML classifiers with transformer-based methods, providing a significant improvement in detection accuracy without an excessively high computational cost.

The system's performance was measured using the new F1 efficiency metric, which we introduced to assess the trade-off between detection accuracy (F1 score) and computational speed. Using this metric, we were able to demonstrate that the proposed system can dynamically adapt to different computational load scenarios, presenting a significant advantage in practical applications.

In conclusion, our study not only provides a comprehensive comparison of SQLi detection methods but also puts forward a novel, efficient approach that could be beneficial in real-world cybersecurity systems. We believe that our contributions lay the groundwork for further research in this area, stimulating the development of even more effective and efficient systems for detecting and mitigating SQLi attacks.

This study examined several types of approaches for SQLi detection in a supervised setting. The high number of parameters of the transformers makes it challenging to be trained on a smaller sample set. It is conjectured that employing self-supervised pre-training on the transformer models would alleviate this problem and help it to capture the nature of SQL queries at high levels more accurately and quickly. This will be investigated in a future study.

\bibliographystyle{IEEEtran}  
\bibliography{references}

\end{document}